\documentclass[prl, superscriptaddress, twocolumn, showpacs, floatfix, nobalancelastpage]{revtex4}

\usepackage{amsfonts}
\usepackage{amssymb}
\usepackage[dvips]{graphicx}
\usepackage{color}
\usepackage{amsmath}

\newcommand{\eend}      {\hspace{\stretch{1}}\rule{1ex}{1ex}}

\begin{document}

\title{Phase Transition in Space: How Far Does a Symmetry Bend Before It Breaks?}

\author{Wojciech H. Zurek}
\affiliation{Theory Division, LANL, MS-B213, Los Alamos, NM  87545, USA}
\author{Uwe Dorner}
\affiliation{Clarendon Laboratory, University of Oxford, Parks Road,
  Oxford OX1 3PU, United Kingdom}

\begin{abstract}
We extend the theory of symmetry breaking dynamics in non-equilibrium second order phase transitions known as the Kibble-Zurek mechanism (KZM) to transitions where the change of phase occurs not in time, but in space. This can be due to a time-independent spatial variation of a field that imposes a phase with one symmetry to the left of where it attains critical value, while allowing spontaneous symmetry breaking to the right of that critical borderline. Topological defects need not form in such a situation. We show, however, that the size, in space, of the ``scar'' over which the order parameter adjusts as it ``bends'' interpolating between the phases with different symmetry follows from a KZM - like approach. As we illustrate on the example of a transverse quantum Ising model, in quantum phase transitions this spatial scale -- the size of the scar -- is directly reflected in the energy spectrum of the system: In particular, it determines the size of the energy gap.

\end{abstract}

\maketitle

\section{Introduction}

Near the critical point of a second order phase transition both the
relaxation time (which determines ``reflexes'' of the system) and the
healing length (which sets the scale on which its order parameter
``heals'' in space, i.e., returns to its equilibrium value) diverge:
\begin{eqnarray}
&&\tau={\tau_0} / {|\epsilon|^{\nu z}},
\label{eq:tau1}\\
&&\xi= {\xi_0} / {|\epsilon|^{\nu}} .
\label{eq:xinu}
\end{eqnarray}
Above, $\epsilon$ is a dimensionless parameter which measures the distance from the critical point. 
For instance, when the transition is caused by the change of the temperature $T$ or by varying
the parameter $g$ of a Hamiltonian, $\epsilon$ is, respectively, given by;
\begin{equation}
\epsilon=\frac {T-T_C} {T_C} , \ \ \ \ \epsilon=\frac {g-g_C} {g_C} \ ,
\label{eq:epsilon1}
\end{equation}
where $T_C$ and $g_C$ are the critical values.

Divergences in $\tau$ and $\xi$ that appear near the critical point as
a consequence of these two equations are often referred to as {\it
  critical slowing down} (for obvious reasons) and as {\it critical
  opalescence} (as the fluctuation on a scale $\xi$ that becomes large
in the vicinity of a critical point cause, in some systems, variations of
the optical properties on scales $\sim\xi$, which leads to
opalescence). These divergences will play a crucial role in our
discussion.

Phase transitions are usually investigated as {\it equilibrium}
phenomena in {\it homogeneous} systems.  But many intriguing questions
arise when a system is driven at a finite pace from one phase to
another.  That this will inevitably happen in the cosmological setting
was first pointed out by Kibble~\cite{Kib76}, who noted that --
because of relativistic causality -- cosmological phase transitions in
a variety of field theoretic models necessarily lead to formation of
topological defects (such as monopoles or cosmic strings) and may have
dramatic astrophysical consequences.

It was later pointed out by one of us \cite{Zur85a} that analogs of cosmological
phase transitions can be studied in the laboratory, and that the {\it
  equilibrium} critical scalings --Eqs.~(\ref{eq:tau1}) and~(\ref{eq:xinu})-- 
can be used to predict various aspects of non-equilibrium dynamics of
symmetry breaking, including the density of topological defects left
behind by a non-equilibrium second order phase transition \cite{Zur85a,
  Zur96a, Kib03}.

The resulting theory (known as ``Kibble-Zurek mechanism'' or KZM) uses
critical scalings of the relaxation time and of the healing length to
estimate the size $\hat \xi$ of the domains that choose the same broken
symmetry \cite{Zur85a, Zur96a}. Owing to the universality of phase
transitions, KZM can be applied on many energy scales, from the
(cosmologically relevant) high energy settings (including experiments
such as RHIC or LHC), all the way to Bose-Einstein condensates at
ultra-low temperatures.

In all of these situations a frequent prediction is that the process
of symmetry breaking will lead to the formation of topological defects.
This is essentially inevitable when the broken symmetry phase
(characterised by the homotopy group) permits their
existence~\cite{Kib76,Zur85a, Zur96a,  Kib03}.  Following the transition, topological
defects should appear with the density of about one defect unit (e.g.,
one monopole or one $\hat \xi$-sized section of a string) per $\hat
\xi$-sized domain. The value of $\hat \xi$ that results from this {\it
  non-equilibrium} process can be deduced from the {\it equilibrium}
near-critical behaviour. Estimating this size is therefore essential.

Here we first summarise the key ideas and equations 
that lead to such estimates. We shall then investigate the behaviour of the
order parameter in an inhomogeneous system, where the change between
phases is imposed by a slowly varying externally controlled parameter,
such as the field in the quantum Ising model. We shall calculate how
far does the transition region between the two phases persist -- how
much does a symmetry that prevails on one side of the critical line
``bend'' before it breaks, spontaneously, on the other side.

\section{Dynamics of Symmetry Breaking}

We consider a second order phase transition that is traversed at a finite rate set by the quench timescale $\tau_Q$,
\begin{equation}
\epsilon=\frac t {\tau_Q} \ .
\label{eq:epst}
\end{equation}
The system will be able to adjust its state adiabatically as long as the rate of change imposed from 
the outside is slow compared to its reaction time given by $\tau$, Eq.~(\ref{eq:tau1}). This change 
from nearly adiabatic to approximately impulse behaviour will happen at an instant $\hat t$
when 
\begin{equation}
\tau(\hat t) = \frac {\epsilon(\hat t)} {\dot \epsilon(\hat t)} \ , 
\quad\text{or}\quad
{\tau_0} {|\frac {\hat t}{\tau_Q}|^{-\nu z}}=\hat t \ .
\label{eq:that2}
\end{equation}
Thus, the state of the system's order parameter will in effect ``freeze'' at
\begin{equation}
\hat t = (\tau_0\tau_Q^{\nu z})^{\frac 1 {1+\nu z}}=\hat \tau .
\label{eq:that3}
\end{equation}
Its evolution will restart only $\hat t$ after the critical point is passed. The instant $\hat t$ plays a
key role in the establishment of the fluctuations which seed structures (such as topological defects)
in the broken symmetry phase \cite{ZZ}. 

The key instant $\hat t$ corresponds to
\begin{equation}
\hat \epsilon = \bigl(\frac {\tau_0} {\tau_Q} \bigr)^{\frac 1 {1+\nu z}}
\label{eq:epshat}
\end{equation}
which in turn sets the characteristic spatial scale given by the corresponding healing length
\begin{equation}
\hat \xi = \xi_0\bigl(\frac {\tau_Q} {\tau_0} \bigr)^{\frac \nu {1+\nu z}}.
\label{eq:xihat}
\end{equation}
This is the estimate of the size of regions that break symmetry in a
more or less coordinated manner~\cite{Zur85a, Zur96a}. Our derivation
subverts equilibrium properties of the system -- its scaling in the
vicinity of the critical point -- to predict non-equilibrium
consequences of the quench. The density of topological defects left behind
in the wake of a phase transition is the best known (but not the only)
example of such predictions of KZM.

These predictions were tested, extended and refined with the help of
numerical simulations \cite{LZ96a, RH00a}, and verified in a variety
of increasingly sophisticated and reliable experiments in liquid
crystals \cite{Chu92a, Bow96a}, superfluids \cite{He4a, He4b, He3},
superconductors \cite{Mon00a, Car00, Man03}, as well as other systems
\cite{Are00}. The majority of the experiments to date are consistent
with KZM. The case of superfluid $^4$He may be an exception.  There
the initial reports of detection of KZM vortices \cite{He4a} were
retracted \cite{He4b} after it turned out that the observed copious
vorticity was inadvertently induced by stirring. It is still not clear
if $^4$He behaviour is really at odds with the KZM predictions
re-evaluated in view of the refined numerical estimates~\cite{RH00a}.
One problem is that the decay of vortex tangle generated by KZM may be
faster than that generated by stirring: KZM leads
to anticorrelations between neighbouring vortices, so phase ordering
kinetics is expected to proceed faster than when vortices are
correlated, as when they originate from turbulent flows. An accessible
recent summary of the experimental situation due to Kibble
\cite{Kib07} is a highly recommended and up-to-date complement to
this brief discussion. Further experiments that may allow for a better
control of various parameters would be obviously welcome.  Gaseous
Bose-Einstein condensates are one recent and very promising proving
ground~\cite{AZ99a,DSBZ02} that has already yielded some exciting
results~\cite{And07a,STC06a,Pfau}.

Recently, KZM theory was applied to quantum phase transitions
\cite{ZDZ05a, Pol05a, Dzi05a, Bodzio1, Levitov, DZ05a, Lam07a, CDDZ07a, DZ07a, Ueda, Cincio, Fubini}. There, the system is at $T=0$, and the nature of its ground state
changes discontinuously as a result of a continuous change of some
parameter of its Hamiltonian. At the critical value of that parameter
the {\it gap} -- energetic price of the lowest excitations, given by
the difference between the energy of the ground and the first excited
state -- is at its minimum, and it disappears in the limit of an
infinite system. In a finite system the quantum phase transition is
then described by an avoided level crossing. The basic observation due
to Damski~\cite{Bodzio1} is that KZM can be used to analyse the avoided
level crossing process, and that it yields results in excellent
agreement with Landau-Zener theory \cite {LZ}. As we shall see
below, the analogy is based on the observation that, far away from the
avoided level crossing, the quantum system usually starts its enforced
evolution adiabatically, as was the case for second order phase
transitions.  However, it can only react on the relaxation timescale
given by the inverse of the gap. Therefore, near the avoided crossing
-- where the gap closes -- its reflexes deteriorate, and the impulse
approximation in the immediate vicinity of a quantum critical point is
appropriate. Below we give an illustration of KZM in action in the exactly solvable
and paradigmatic quantum Ising model.

\section{Quench in a Quantum Ising Model}

According to Sachdev \cite{Sac99a} the {\it quantum Ising model} is
one of two prototypical examples of quantum phase transitions.
It represents a chain of spins with the Hamiltonian
\begin{equation}
H=-J(t) \sum_{l=1}^N \sigma_l^x - W\sum_{l=1}^{N-1} \sigma_l^z \sigma_{l+1}^z  
\label{eq:Hamiltonian}
\end{equation}
where $\sigma_l^{x,z}$ are Pauli operators, $W$ is the Ising coupling,
and $J(t)$ is due to the external (e.g., magnetic) field that attempts
to align all spins with the $x$-axis. The phase transition from the
paramagnetic state where all the spins are aligned with $x$ by the
strong external field (e.g., $|\rightarrow,\rightarrow, \dots,
\rightarrow \rangle$) to the low-field ferromagnetic and degenerate
(in the large $N$ limit) ground state that ``lives'' in the Hilbert
subspace spanned by the broken symmetry basis $|\uparrow, \uparrow,
\dots, \uparrow\rangle$ and $ |\downarrow, \downarrow, \dots
\downarrow\rangle$ takes place when $J(t)=W$. Therefore, as suggested by
Eqs.~(\ref{eq:epsilon1} and~\ref{eq:epst}), a key role is played by the {\it relative coupling} given by
\begin{equation}
\epsilon(t)={J(t) / W} - 1 = t/\tau_Q
\label{eq:epsilon}
\end{equation}
and we assume that it depends linearly on $t$.

All the relevant properties depend on the size of the {\it gap}
$\Delta$ between the ground state and the first excited state. In an
infinite system the gap is given by:
\begin{equation}
\Delta=2|J(t)-W|=2W|\epsilon(t)| \ .
\label{eq:gap}
\end{equation} 
The spectrum (and the gap) of a finite quantum Ising system are
illustrated in Fig.~1. The gap sets the energy scale and is reflected
in the {\it relaxation time} and the {\it healing length}
\begin{eqnarray}
&&\tau={\hbar /  \Delta} = {\hbar / {2 W| \epsilon(t)| }}={\tau_0 / |\epsilon(t)| } \ ,
\label{eq:tau} \\
&&\xi=2Wa/\Delta(t) = a / |\epsilon(t)| = \xi_0/|\epsilon(t)| \ ,
\label{eq:xi}
\end{eqnarray}
the latter given by the product of $\tau$ and the speed of sound $c=2Wa/\hbar$ (see \cite{Sac99a}), where $a$ is the distance
between spins. The divergence of $\tau$ and $\xi$ is the critical
slowing down and the analogue of critical opalescence, respectively.

The instants $\pm \hat t$ when the behaviour switches from adiabatic to impulse happen when the reaction time of the system, 
Eq.~(\ref{eq:tau}), is the same as the timescale on which its Hamiltonian 
is altered (given by $\epsilon(t)/\dot\epsilon(t)=t$), i.e.
\begin{equation}
\tau(\hat t) = \ {\tau_0 / | \epsilon(\hat t)| } \  = \  {{ \epsilon(\hat t)} / {\dot \epsilon(\hat t)}} = \hat t,
\label{eq:thateq}
\end{equation}
and thus
\begin{equation}
\hat t = \sqrt {\tau_Q \tau_0} = \sqrt{{\tau_Q \hbar} / {2 W}}.
\label{eq:that}
\end{equation}
As before, for $t<-\hat t$, the state of the system will continue to
adjust adiabatically to changes imposed by the decreasing $J(t)$.
However, at $t= - \hat t$ before the critical point the evolution will
cease, and it will re-start only at $t=+\hat t$ after the transition,
with the initial state similar to the one ``frozen out'' at $-\hat t$.

Using the relative coupling $\hat \epsilon$ associated with $\hat t$ we get
\begin{eqnarray}
&& \hat \epsilon  \equiv   \epsilon(\hat t)  ={ {\hat t} /  \tau_Q} = \sqrt{ \tau_0 /  \tau_Q},
\label{eq:epshat1}\\
&& \hat \xi  \equiv \xi_0/{\hat \epsilon} =  \xi_0 \sqrt{ \tau_Q / \tau_0} = a \sqrt{{2W \tau_Q} / {\hbar}} \ .
\label{eq:xihat1}
\end{eqnarray}
Following KZM, we now predict appearance of $O(1)$ defects per $\hat \xi$.  
Their density should be approximately
\begin{equation}
\hat n_{KZM} \simeq  a /  \hat \xi =  \sqrt{ {\hbar}  /  {2W \tau_Q}  }
\label{eq:Nhat}
\end{equation}
per spin. This is only an estimate: Simulations of classical second
order phase transitions yield defect densities that scale with
$\tau_Q$ in accordance with this reasoning, but are lower than the
relevant power of $\hat \xi$ of Eq. (\ref{eq:xihat1}) so that a ``unit
of defect'' is often separated by $\sim$10-15 $\hat \xi$; see e.g.
\cite{LZ96a, RH00a}.

\begin{figure}[tp]
\centering  
\includegraphics{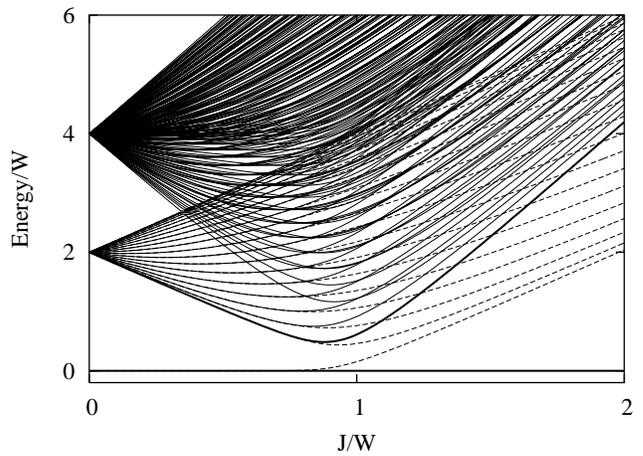}
\caption{%
  Energies of lowest excitations of the Ising chain for $N=20$ with open
  (rather than periodic) boundary conditions. The ground and the first
  accessible excited state that define the gap are marked with a
  thicker line.}
\label{spectrum}
\end{figure} 

There is a fundamental difference between quantum and thermodynamic
phase transitions.  In the thermodynamic case `real' fluctuations
exist above the critical point. They can initiate the symmetry breaking
process -- in effect, choose how symmetry breaks in domains that
appear after the transition. In the case of quantum phase transition 
there are only  `quantum fluctuations', but they are virtual, so one
cannot be certain that they will have an analogous effect on the
post-transition state. Therefore, an explicitly quantum approach to
the quantum Ising model is needed.

As we have already noted, a key feature of a quantum phase transition
is the gap $\Delta$. Actually, as can be seen in Fig.~\ref{spectrum},
the relevant gap (i.e. the gap between the ground and the first
accessible state) is not the symmetric $\Delta$ but rather $2\Delta$
on the approaching side. Nevertheless, in the quantum Ising model the
gap disappears at the critical point when the system is infinite.
When $N < \infty$, this critical gap becomes small, but does not
disappear, (see Fig.~\ref{spectrum}). This is important, and allows us
to propose a purely quantum approach: Instead of density of defects in
an infinite system we can compute (as a function of quench timescale
$\tau_Q$) the length of the spin chain that, with probability $p$ of
about a half, can remain defect - free (in a ground state) after the
quench. An excited state would (most likely) contain just a single
excitation (e.g., a kink). Therefore, the {\it inverse} of the length
that limits the excitation probability to approximately $\sim 0.5$
should correspond to about half of the kink density.

The two lowest accessible levels of $H$ in the vicinity of the
critical point (Fig.~\ref{spectrum}) exhibit an avoided level crossing
and we can calculate the excitation probability of the system driven
through this avoided crossing using the Landau-Zener formula (LZF)~\cite{LZ},
\begin{equation}
p \simeq e^{ -\frac {\pi \hat \Delta^2}{2 \hbar |v|}} \ .
\label{eq:LZF}
\end{equation}
Here $\hat \Delta$ is the minimum gap between the two levels and $v$
is the velocity with which the transition is imposed on the system
given by $v=\dot \Delta$ far away from the avoided crossing. The KZM
approach was shown by Damski to provide an excellent approximation to
LZF \cite{Bodzio1} (see also \cite{DZ05a} for extensions). Using LZF
we compute the size $\tilde N$ of the spin chain that will probably 
remain in the ground state in course of the quench with probability
$p\sim 0.5$.  Equation (\ref{eq:LZF}) translates into a condition for
the rate of quench that produces a kink with probability $p$,
\begin{equation}
|v| \leq \frac{\pi \hat \Delta^2}{2 \hbar |\ln p|} \ .
\label{eq:speed}
\end{equation}
Using $v=|\dot \Delta| = 2\dot J(t)={2W/ \tau_Q}$ [see Eq.~(\ref{eq:epsilon})] and $\hat \Delta = 4 \pi W / N $ for the gap upon ``closest approach'' we get
\begin{equation}
|v| =|\dot \Delta| = \frac{2 W}{\tau_Q}  \leq \frac{\pi (4 \pi W/ \tilde N)^2}{2 \hbar |\ln p|} \ .
\end{equation}
This relates the size $\tilde N$ of a chain that will remain defect-free with the probability 1-$p$ 
to the quench rate:
\begin{equation}
\tilde N \leq  2 \pi \sqrt{ \frac{\pi W \tau_Q}{\hbar |\ln p|} }.
\label{eq:ntilde}
\end{equation}
This LZF estimate is (surprisingly) accurate for $p<0.5$ even though
there are many levels in the spectrum of the quantum Ising model. Such
accuracy was surprising when this analysis was first
presented~\cite{ZDZ05a}. Dziarmaga~\cite{Dzi05a} has soon after
demonstrated that this is no accident -- in effect, the phase
transition in the quantum Ising model can be decomposed into a collection
of independent avoided level crossings.

We can now directly compare KZM, Eq. (\ref{eq:Nhat}), and LZF
predictions for defect density:
\begin{equation}
\tilde n_{LZF} \simeq \frac{1}{\tilde N} = \frac{1}{2 \pi}  \sqrt{\frac{2 |\ln p|}{\pi} } \times \hat n_{KZM}.
\label{eq:ntildehat}
\end{equation}
The two estimates exhibit the same scaling with the quench rate and
with the parameters of $H$, Eq.~(\ref{eq:Hamiltonian}).
\begin{figure}[tp]
\centering  
\includegraphics{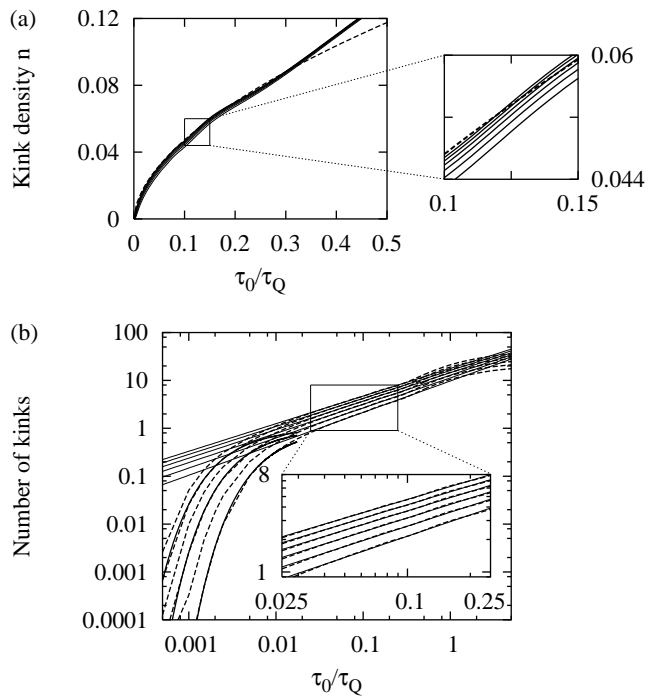}
\caption{{\bf (a)} Number of kinks per spin in the quantum Ising model after a quench starting 
  in the ground state at $J=5W$ and ending at $J=0$, plotted as a
  function of the quench rate $\tau_0/\tau_Q = \hbar v/4W^2$ for
  $N=50,60,70,80,90,100$ (solid lines; bottom to top). The scaling
  $\hat n_{KZM} \sim \sqrt {\tau_0 / \tau_Q}$ predicted by KZM is
  consistent with the simulations (see \cite{Dor03} for details of the
  numerical method).  Agreement improves with the size of the system:
  for 100 spins a fit gives $n \sim \tau_Q^{-0.58}$ (dashed line).  As
  in the classical case~\cite{LZ96a} Eq. (\ref{eq:Nhat}) is an
  overestimate; the best fit is $n \simeq 0.16 \hat n_{KZM}$. {\bf
    (b)} Total number of kinks for $N=50,60,70,80,90,100$ spins
  (dashed lines, bottom to top) after a quench as a function of the
  quench rate $\tau_0/\tau_Q = \hbar v/4W^2$.  Both the scaling $\hat
  n_{KZM} \sim 1 / \sqrt \tau_Q$ predicted by KZM (solid, straight
  lines), Eq.~(\ref{eq:Nhat}), and the LZF estimate $p$ (solid, bend
  lines) when less than one kink is expected are valid. The straight
  lines are linear fits in the range (0.025,0.25) yielding slopes
  between 0.66 and 0.58.  Numerical data include these used in (a) but
  we now go beyond the expected range of validity of KZM. For
  sufficiently slow quenches LZF provides reliable predictions. Very
  fast quenches are ``all impulse'', levelling off of the expected
  number of kinks, as is indeed seen.}
\label{lzfkzm}
\end{figure} 
LZF predicts fewer defects than ``raw KZM'' ($\tilde n_{LZF} \simeq
0.14 \times \hat n_{KZM}$ when $p$ is set -- somewhat arbitrarily --
to 0.5). This is not a big surprise -- as seen in the numerical simulations,
confirmed by the experiments and verified analytically in specific
models, Eqs.~(\ref{eq:xihat1}) and (\ref{eq:Nhat}) provide correct
scalings, but tend to overestimate densities (see e.g. \cite{LZ96a, RH00a}). 
Fig.~\ref{lzfkzm} indicates that this conclusion holds also
for the quantum Ising model.

Figure~\ref{lzfkzm} also shows that the kink density scales
approximately as $ \sim 1/\sqrt{\tau_Q}$, Eq. (\ref{eq:Nhat}), in the
region of the validity of KZM, i.e. for $\hat \epsilon$ less than 1
(so that quench is quasi - adiabatic at the beginning and at the end,
but impulse near the critical point, i.e., when at least one defect is
expected). The prefactor which is approximately $0.16$ [0.12 if the steeper slope on the
approach in Fig.~\ref{spectrum} is taken] is not far from the previous
experience \cite{LZ96a}. For very slow quenches ($\hat t > \hbar /
{\hat \Delta}$, or $\tau_Q > (\frac{N}{2 \pi})^2 \tau_0$), i.e. if the
system is nowhere convincingly `impulse', LZF is surprisingly
accurate. We conclude that the two approaches work well in
complementary regimes of quench rates, and predict the same scaling of
the size of broken symmetry domains with quench time.

The importance of the behaviour of the gap (and, in particular, of the
smallest gap $\hat \Delta$) for the quantum estimate of defect density
is -- in addition to the broader theme of connections between KZM and
LZF -- our principal general conclusion at this point. We shall return to discuss
the gap in the quantum Ising model and utility of extensions of KZM and LZF 
to more complicated behaviours of the near-critical gap below.

\section{Statics of symmetry breaking}

Consider now a situation where the all-important dimensionless
$\epsilon$ is {\it time-independent}, but depends on the location in
space instead. We suppose -- in analogy with
Eqs.~(\ref{eq:epst},~\ref{eq:epsilon}) -- that in the vicinity of the
critical point $\epsilon(x)$ is approximately linear:
\begin{equation}
\epsilon=\frac x {\lambda_Q} = \alpha x \ .
\label{eq:epsx}
\end{equation}
Clearly, sufficiently far from $x=0$ the order parameter will settle
into an equilibrium state corresponding to the local value of
$\epsilon$. We are however left with an interesting question: How far
-- and how -- does the influence of the critical point propagate?

There is an intriguing analogy between the freeze-out in time we have
discussed in the preceding section and the freeze-out in space we are
led to consider here: When $\epsilon$ changes in space slowly compared
to the local healing length given by $\xi={\xi_0} / {|\epsilon|^{\nu}}
$, Eq.~(\ref{eq:xinu}), the order parameter will be able to adjust in
space to the changes of $\epsilon$ -- i.e., there should be a
well-defined local value of the order parameter, local healing length
$\xi$, and local relaxation time $\tau$, etc. However, very close to
$x=0$ the system cannot ``heal'' fast enough: The healing length
becomes large, much larger than $x$. Consequently, the critical
opalescence -- the divergence of $\xi= {\xi_0} / {|\epsilon|^{\nu}}$
-- will open up a ``scar'' in the order parameter that does not
properly heal.  Our aim now is to describe the consequences. 
To find out the size of the scar we write down the spatial analogue of
Eq.~(\ref{eq:that2}),
\begin{equation}
\xi(\hat x) = \frac {\epsilon(x)} {\partial_x \epsilon(x) }|_{x=\hat x} \ , 
\label{eq:xhat1}
\end{equation}
for the ``adiabatic-impulse'' borderline point, $\hat x$. Thus,
\begin{equation}
{\xi_0} {|\frac {\hat x}{\lambda_Q}|^{-\nu}}=\hat x \ .
\label{eq:xhat2}
\end{equation}
Consequently, and in accord with previous discussion, we conclude that the order parameter will 
in effect ``freeze'' (or, to put it differently, that the scar opened up by the transition through the critical region will heal) at a distance
\begin{equation}
\hat x = (\xi_0\lambda_Q^{\nu})^{\frac 1 {1+\nu}}=\hat \xi 
\label{eq:xhat3}
\end{equation}
from the critical point. The distance $\hat x$ corresponds to
\begin{equation}
\hat \epsilon = \bigl(\frac {\xi_0} {\lambda_Q} \bigr)^{\frac 1 {1+\nu}}.
\label{eq:epsxhat}
\end{equation}
We are really done -- there is just a spatial $x$, rather than both $x$ and $t$, as before. But let 
us (as a consistency check) repeat and take a few more steps, and use this estimate of $\hat \epsilon$ to calculate the characteristic spatial scale given by the corresponding healing length,
\begin{equation}
\hat \xi = \xi_0\bigl(\frac {\lambda_Q} {\xi_0} \bigr)^{\frac \nu {1+\nu }}.
\label{eq:xxihat}
\end{equation}
We know this already [see Eq.~(\ref{eq:xhat3})]. This is
now the estimate of the size of the scar -- the size of the region
that is ``still thinking" about how to break symmetry. Earlier, our
derivation of the freezeout time and of the resulting analysis of the
dynamics of symmetry breaking subverted equilibrium properties of the
system -- its scaling in the vicinity of the critical point -- to
predict non-equilibrium consequences of the quench. We have now
repeated this subversive strategy in a new setting, by using the healing
length (that is, strictly speaking, defined in a {\it homogeneous}
system) to find out consequences of inhomogeneity that must be there
in the vicinity of the critical point if the phase transition takes
place in space. The structure of the scar in the order parameter that
connects the two phases is now of interest {\it per se}.

There are two specific values of $\nu$ that are often encountered: As
we have seen in the preceding section, for quantum Ising model
$\nu=1$. Therefore,
\begin{equation}
\hat x = \sqrt {\xi_0\lambda_Q} =\hat \xi 
\label{eq:xhatIsing}
\end{equation}
and the size of the scar is simply a geometric average of the two relevant lengths in the problem.
We shall see below that, as a consequence, in the quantum Ising model the presence of such a scar 
widens a critical gap.
In a mean field theory we have $\nu=1/2$. Consequently,
\begin{equation}
\hat x = (\xi_0^2\lambda_Q)^{\frac 1 3 }=\hat \xi.
\label{eq:xhatmean}
\end{equation}
The thickness of the boundary between the two layers is a compromise that involves the characteristic
healing length of the order parameter on one hand, and the typical scale of the externally imposed 
inhomogeneity of the Hamiltonian on the other. The extent to which each of them has its say depends
on the scaling exponent $\nu$. The most obvious test of these predictions would then explore variations 
of the order parameter in the vicinity of an externally imposed inhomogeneity. 

\section{Opening a gap with a wedge of a field}

\begin{figure}[tp]
\centering  
\includegraphics{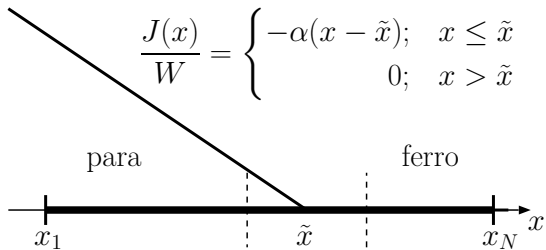}
\caption{%
  Illustration of the spatially dependent field $J(x)$ in the
  transverse quantum Ising model we consider. The parameter $J(x)$
  corresponds to $J(t)$ from Hamiltonian~(\ref{eq:Hamiltonian}) but
  now depends on position, $J(t) \rightarrow J(x)$. The spins are
  located at positions $x_l = (l-1)a$ with $l=1,\ldots,N$.}
\label{slant}
\end{figure}

To illustrate our considerations with a concrete example we return to
the quantum Ising model we have already employed in the previous section.
The Hamiltonian we shall use now has the same form as before, Eq.
(\ref{eq:Hamiltonian}), but the field $J$, which before was the same for
every spin, now has a form illustrated in Fig.~\ref{slant}. As a
result, sufficiently far to the left of $\tilde x$ the system will be
in a paramagnetic phase, while to the right a ferromagnetic phase should
be dominant.

Our task is to characterise the state of such a system. In a homogeneous quantum 
Ising spin system the correlation function given by 
\begin{equation}
\zeta(k)=\langle  \sigma_{l-k}^z \sigma_{l}^z \rangle 
\label{eq:corr}
\end{equation}
is a well defined object, whose behaviour is thoroughly
explored~\cite{Sac99a}. Indeed, the decay of such correlation
functions is used to define the coherence length (which is in effect
often equal to the healing length we have employed earlier). In a
homogeneous system $\langle \sigma_{l-k}^z \sigma_{l}^z \rangle$
obviously does not depend on the ``reference spin'' $l$: Each spin
``lives'' in an identical neighbourhood, so $\zeta(k)$ is
translationally invariant with respect to $l$. But when a system is
inhomogeneous -- for example when a phase transition occurs at some point
within the system -- the correlation function will obviously depend on
where the reference spin $l$ is. We therefore consider a conditional
correlation function,
\begin{equation}
\mu(k|l)=\langle  \sigma_{l-k}^z \sigma_{l}^z \rangle.
\label{eq:correl}
\end{equation}
It depends explicitly on the location of the reference spin $l$.
Formally, $\zeta(k)$ is given by $\mu(k|l)$ averaged over all $l$. In
a homogeneous case $\mu(k|l)$ is independent of $l$, so this
averaging is trivial. On the other hand, in an inhomogeneous system we
are investigating here, such averaging would be counterproductive: 
The averaging would obscure precisely the signature we are looking for,
i.e., the imprint made by the inhomogeneity of the Hamiltonian on the
state of the system.
\begin{figure}[tp]
\centering  
\includegraphics{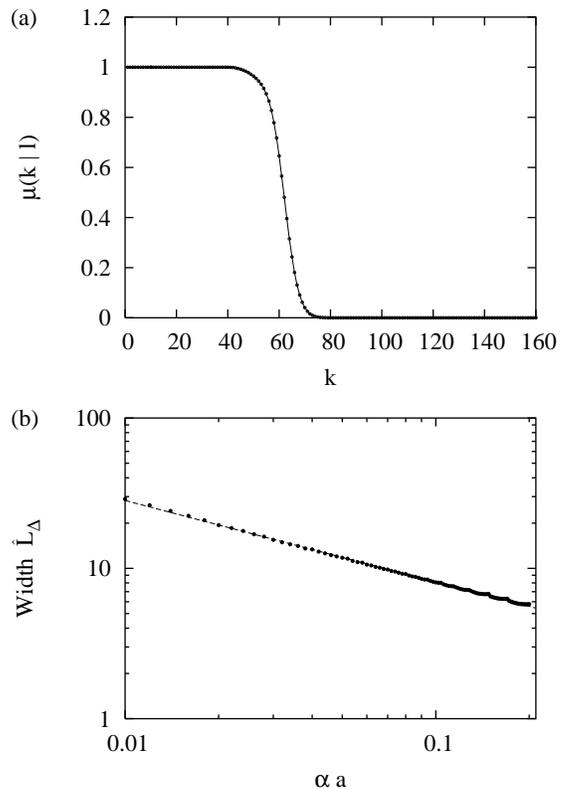}
\caption{%
  (a) Conditional correlation function, $\mu(k|l)$, for $N=200$,
  $\tilde x/a = 120$, $\alpha a=0.05$ and $l=160$. The reference spin
  located at $x=x_{l}$ is in the ferromagnetic region. (b) Width of
  the transition region of the conditional correlation function
  $\mu(k|l)$ versus $\alpha$ for $N=200$, $\tilde x/a = 120$ and $l=160$.
  The dashed line is a fit to the data leading to $\hat L_\Delta =
  2.34(\alpha a)^{-0.54} = 2.34(\lambda_Q/a)^{0.54}$.}
\label{corrfunction}
\end{figure} 

In Fig.~\ref{corrfunction}a we show $\mu(k|l)$ for the fixed reference
spin $l$ located in the ferromagnetic region.  The behaviour of
$\mu(k|l)$ seen there is a good illustration of our previous
discussion: For spins located in the ferromagnetic region
$\mu(k|l)\approx 1$. However, when $k$ assumes values such that $l-k$
falls within the paramagnetic region, the correlation function
vanishes, $\mu(k|l)\approx 0$. The transition between these two
extremes is our focus of interest. We can characterise the size of the
``scar'' by quantifying the width of the transition regime. The most
obvious way to do this is through the inverse of the slope of
$\mu(k|l)$ where $\mu(k|l) = \frac{1}{2}$. The resulting width of the
transition region is plotted in Fig.~\ref{corrfunction}b. Evidently,
the behaviour predicted by the KZM-like discussion for the quantum
Ising model, Eq.~(\ref{eq:xhatIsing}), is reflected in the
correlations between spins. Such correlation functions should be
experimentally accessible, and may be a good way to test our
predictions.

We note that some of the effects of the inability of the order
parameter to adjust to the inhomogeneous variations of the Hamiltonian
may become apparent even if the system does not cross ``all the way''
into the other phase. For instance, just coming close to the critical
point may lead to the structures with sizes that can be estimated
using the above approach. Similarly, when the same phase is separated
by a narrow strip when the parameter that controls its phase crosses
the critical point (i.e., a strip in space within which $\epsilon$ has
a different sign than outside it), an extension of the above
discussion should be applicable. There should be therefore lots of
opportunities to experimentally test the above equations, and plenty
of variations of the homogeneity - inhomogeneity theme. One such obvious
variation involves adopting a nonlinear spatial dependence of
$\epsilon$ on $x$ -- i.e., a dependence that is different from
Eq.~(\ref{eq:epsx}). But there are clearly many more. We shall not
attempt to enumerate them here.

\begin{figure}[tp]
\centering  
\includegraphics{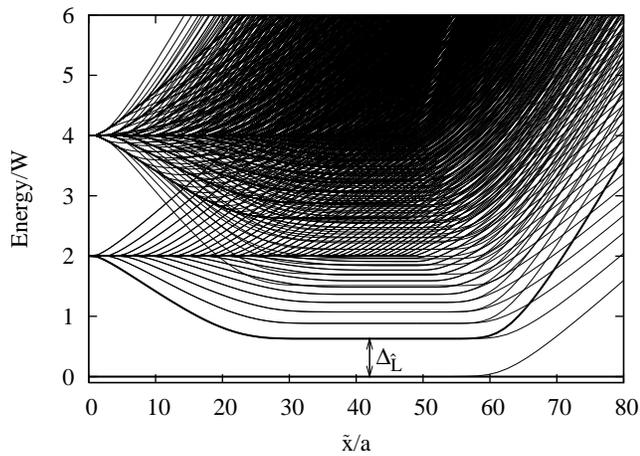}
\caption{%
  Energies of lowest excitations for $N=50,\,\alpha=0.05/a$ versus $\tilde x$. The
  ground and the first accessible excited state that define the gap
  are marked with a thicker line.}
\label{spectrum_inh}
\end{figure}

Let us instead point out a less obvious consequence of a spatial
inhomogeneity that will be especially important in quantum phase
transitions: Its effect on the gap.  The spectrum of an inhomogeneous
Ising system as a function of the location of the point $\tilde x$ --
the place where $J(x)$ becomes 0, see Fig.~\ref{slant} -- is shown in
Fig.~\ref{spectrum_inh}. When this figure is compared with
Fig.~\ref{spectrum}, the spectrum of a homogeneous system, two
striking differences in the behaviour of the gap emerge. (i) Instead of
a sharp ``corner'' where $J/W=1$, there is now a plateau
which extends over a range of values of $\tilde x$. (ii) Moreover,
this plateau is lifted above where it was (at $\sim 4 \pi W / N$) for
a chain of $N$ spins. In short, the sharp gap minimum we had before becomes
now a wider, extended ``bottleneck''.
The spatial structure induced by the inhomogeneity in the state of the
system should have an effect on their energetics. Indeed, one could
venture a guess (based on ``general principles'') that the size of the
gap -- which is the property of the whole inhomogeneous spin chain --
will change with the size of the scar, $\hat x$. One could moreover
speculate (and here the ``general principles'' are becoming too
general for comfort) that the size of the gap should be inversely
proportional to (some power of) $\hat x$. We shall now pose this very
same question more precisely, and arrive at the estimate of the basic
parameters of the ``bottleneck'' more directly.

The inhomogeneous Ising system we are dealing with can be typically
divided into three regions: A {\it ferromagnetic domain} (where the
Ising coupling $W$ dominates), a {\it paramagnetic domain} (where
$J(x)$ is large) and what's in between these two -- the {\it
  near-critical domain}, where $|J(x)-W|$ is small.  The energetic
price of excitations in either ferromagnetic or paramagnetic domains
is large compared to the size of the gap, which (we remind the reader)
in a near-critical chain of $L$ spins is given by $\Delta_L\simeq 4
\pi W / L$. 

\begin{figure}[tp]
\centering  
\includegraphics[width=7cm]{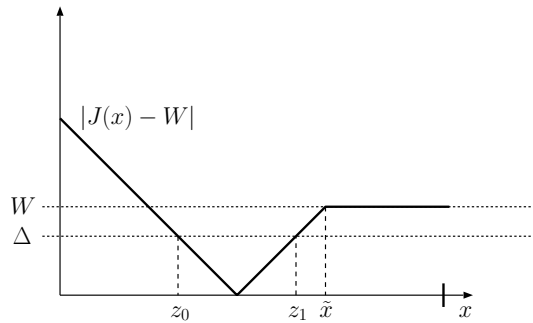}
\caption{Illustration of inequality~(\ref{eq:ingap}) which defines a critical region of length $\hat X_\Delta = z_1-z_0 = 2\Delta/\alpha W$.}
\label{slant2}
\end{figure}

Our aim is to compute the size of the bottleneck gap in the
inhomogeneous case of Fig.~\ref{slant}.  We shall do that by a
self-consistent estimate suggested by the above division of the whole
inhomogeneous system into three domains. We first note that the
energetic price of the excitations outside the near-critical regime is
prohibitively large whenever these regimes are well defined, so such
excitations will not be relevant for the calculation of the size of
the gap (which is defined as the ``price'' of the least energetically
expensive excitation). On the other hand, the near-critical domain
should be able to support collective excitations that -- when the size
of that domain is $L$ -- should be able to support collective
eigenstates with energies given by $\Delta_L\simeq 4 \pi W / L$. Thus, 
all we have to do is to find that size $L$. One might be at this
stage tempted to venture a guess that $L \simeq \hat x$, but let us
proceed cautiously, and follow a line of reasoning based on energetics
we have outlined above. To this end we propose a self-consistency
condition,
\begin{equation}
|J(x)-W|<\Delta ,
\label{eq:ingap}
\end{equation}
which in effect defines the near-critical domain. A plot that illustrates this inequality is shown in Fig.~\ref{slant2}.

\begin{figure}[tp]
\centering  
\includegraphics{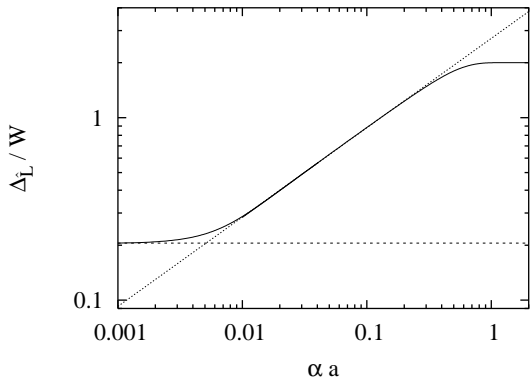}
\caption{%
  The size of the smallest gap between the ground state and the first
  accessible excited state -- the level of the plateau in Fig.
  \ref{spectrum_inh} -- as a function of the slope
  $\alpha=1/\lambda_Q$ for $N=50$ spins. The dotted line is a fit to
  the data between $\alpha a=0.01$ and $\alpha a = 0.2$ leading to
  $\Delta_{\hat L}/W = 2.72(\alpha a)^{0.49}$. The horizontal, dashed
  line corresponds to the minimum gap of a homogeneous spin chain.}
\label{minimum_gap}
\end{figure}

At this point the relevant $\Delta$ is not known. But (as $|J(x)-W|$
is small in the near-critical domain) it is natural to employ
$\Delta_L\sim 4 \pi W / L$, where now $L$ stands for the number of
spins that contribute to the collective state that defines the gap, i.e., the spins 
that fit inside the near-critical domain. Using $J(x)/W=-\alpha (x-
\tilde x)$ of Fig.~\ref{slant}, we solve the above equation to
obtain the extent, in space, of the near-critical domain that defines
the size of the gap,
\begin{equation}
\hat X_{\Delta} = \sqrt{ {8\pi a}/{\alpha}} \approx 5.01 \sqrt {{a}/{\alpha}},
\label{eq:Xgap}
\end{equation}
or, in number of spins (rather than in distance),
\begin{equation}
\hat L_{\Delta}= \hat X_{\Delta}/a = \sqrt{ {8\pi}/{\alpha a}} \approx 5.01 /\sqrt{\alpha a}.
\label{eq:Lgap}
\end{equation}
This immediately yields an estimate of the gap size,
\begin{equation}
\Delta_{\hat L} = 4 \pi W / {\hat L_{\Delta}} \approx 2.51 \sqrt{\alpha a}W.
\label{eq:hatlgap}
\end{equation}
With the help of these results we now understand the basic structure
of the eigenenergies, and, in particular, of the gap seen in
Fig.~\ref{spectrum_inh}. Figure~\ref{minimum_gap} shows the size of
the gap as a function of the slant $\alpha=1/\lambda_Q$. The square root dependence predicted by Eq.~(\ref{eq:hatlgap}) is
evident. Indeed, even the prefactor obtained through our simple
estimate above is close to the one that is obtained from the numerical
experiment. There are obvious limits to the validity of our derivation
that are also in evidence with Fig.~\ref{minimum_gap}: When $\alpha$
is so small that the system as a whole is approximately homogeneous,
the near-critical domain extends over the whole system, $\sqrt{ \frac
  {8\pi}{\alpha a}}=\hat L_{\Delta}$ is comparable to, or larger
than $N$, and the asymptotic size of the gap $\sim 4 \pi W / N$ for
the homogeneous system is attained. On the other hand, when $\alpha$
is so large that the size of the scar becomes comparable with $a$, the
spacing between the spins, the near-critical region disappears
altogether, and the size of the gap is set by $2W$.

\section{Discussion}

We have presented an approach to symmetry breaking phase transitions
that occur in space.  As in the original KZM (which employs
equilibrium near-critical scalings of the relaxation time $\tau$ and
of the healing length $\xi$), it rests on the near-critical behaviour
of $\xi$. While the healing length is defined in a homogeneous system,
we have employed its variation with $\epsilon$ (dimensionless
parameter that measures the distance from the critical point) to
predict what happens near a critical point that occurs in space in an
obviously inhomogeneous system. The resulting theory yields the size
of the scar of the transition region between the two phases.

We have applied these results to the quantum Ising model. There the
gradient of the external field leads to a gradual re-alignment of the
spins. The border between the paramagnetic and ferromagnetic phases
has a structure that can be characterised by the correlations between
spins, which change between the two patterns of alignment. We have
seen that this change is gradual, and happens (as was predicted by our
theory) on the scale inversely proportional to the square root of the
gradient of the field. The simple theory leading to Eq.
(\ref{eq:xxihat}) results in correct scaling behaviour. As expected, a
more specific calculation based on the quantum Ising model leads to a
different prefactor, Eqs. (\ref{eq:Xgap},~\ref{eq:Lgap}). In effect,
we now have two different measures of the size of the scar, one based
on the correlations, Fig. \ref{corrfunction}, and the other on the
size of the gap. We do not know which one of them (if any) is
``correct'', but we point out that they can differ.

Existence of the ``transition scar'' has a dramatic effect on the
eigenspectrum of the system. In particular, instead of the simple
avoided level crossing where the minimum gap helps set the probability
that the system will remain in its ground state we are now dealing
with a an extended ``bottleneck gap''.  Parameters of this bottleneck
gap are (including a prefactor, somewhat in contrast with the case of
the scar size estimates) in a surprisingly good accord with the
theory.

One is now tempted to consider quantum phase transitions that occur
both in space and in time. For thermodynamic phase transitions this
situation can be treated by the KZM-based approach of Kibble and
Volovik \cite{KV}. Their conclusion was, in effect, that when the
critical line moves through space with velocities in excess of $\hat v
= \hat \xi / \hat \tau$, the quench that leads to spontaneous symmetry
breaking will proceed as if it were homogeneous, and KZM estimates of
defect densities relevant for homogeneous quenches will apply. On the
other hand, in the limit of slowly propagating phase fronts defect
production will be suppressed, as the symmetry breaking choices made
by the order parameter will be simply propagated with the phase front.
This conclusion seems to be only weakly influenced by the spatial
gradient of $\epsilon$, and is now supported by numerical studies (see
e.g. \cite{DLZ}).

Our preliminary numerical simulations of the same problem in the
quantum case indicate that there the situation is more complicated.
Part of the problem can be traced to the fact that the dynamics of a
quantum phase transition is strictly reversible. Thus, the energy
deposited in the order parameter stays in the order parameter, and --
e.g., in the quantum Ising model -- has ``nowhere to go'' except into
the kinks. As a result, the rate of creation of kinks now depends on
the spatial gradient of $\epsilon$, and does not become suppressed as
dramatically as before when the front velocity falls below $\hat v$.
\begin{figure}[tp]
\centering  
\includegraphics{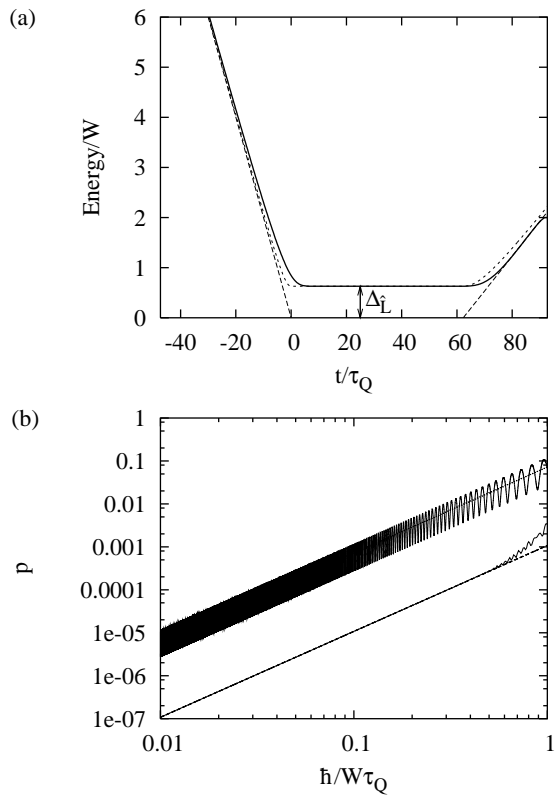}
\caption{%
  (a) Energy of the first accessible state for $N=80$,
  $\alpha=0.05/a$ (solid line). The slant is moving with a constant
  velocity $a/\tau_Q$ across the spin chain, i.e. $\tilde x(t) = x_0 -
  at/\tau_Q$. $x_0$ is chosen such that the flat part of the spectrum
  begins at $t=0$. The dashed lines are linear fits to the solid line
  in order to get parameters for a Landau-Zener type two-level
  Hamiltonian the excited state of which is given by the dotted line.
  (b) Upper line: Solution of a Landau-Zener type problem with an
  energy spectrum given by the dotted line in (a). The function
  oscillates rapidly around the value
  $7.2\times10^{-2}(\hbar/W\tau_Q)^2$ (dotted line). The lower line
  shows the excitation probability for $N=80$, $\alpha=0.05/a$. The
  dashed line is a fit leading to
  $1.1\times10^{-3}(\hbar/W\tau_Q)^{2.0}$.}
\label{LZ}
\end{figure} 
Moreover, in quantum phase transitions, the limit when less than a
single kink is expected in the whole system is explicitly quantum, and
(as we have seen earlier) different than (although compatible with)
the KZM scaling. However, in an inhomogeneous case the simple
Landau-Zener formula (suitable for gaps of Fig. 1) is obviously
inapplicable. We have tried out its natural generalisation (which
approximates the spectrum of Fig.~\ref{spectrum_inh} with two obvious
avoided level crossings connected with a plateau (see Fig.~\ref{LZ}a).
The fit is imperfect. The resulting prediction of the probability of a
kink, Fig.~\ref{LZ}b, is also not very satisfying. But this
calculation (while it predicts spurious oscillations, and
overestimates the rate by about an order of magnitude) does capture
one essential feature: The dependence of the probability $p$ on the
quench rate $1/\tau_Q$ is now no longer exponential, as it was in the
original LZF, but it becomes quadratic. Such quadratic dependence
appears automatically, if rather surprisingly -- given the exponential
nature of LZF, Eq. (\ref{eq:LZF}) -- in the avoided level crossing
problem when the transition starts at the place of the nearest
approach of the two levels~\cite{DZ05a,CDDZ07a}, rather than far away from the
avoided crossing, as is usually assumed~\cite{LZ}. The behaviour we
observe may be due to the asymmetry of the gap.  In effect, as the
slanted potential of Fig.~\ref{slant} traverses the spin chain, the
system will generally approach the plateau of the bottleneck gap with
a slope that is quite different from the slope on the other end
of the gap, where it exits.  Therefore, recent study ~\cite{DZ05a} --
showing that when avoided level crossing is traversed starting at the
minimum of the gap, the probability of excitation $p$ on the quench
rate $1/\tau_Q$ is no longer exponential but that it becomes quadratic
-- becomes relevant.  For instance, in the case of the asymmetric gap
of Fig.~\ref{LZ}a, one can imagine that the system is delivered
essentially in its ground state to the bottleneck of the gap. While
$\tilde x$ moves along the flat part of the gap, the Hamiltonian of
the system is unchanged, so no transitions happen.  However, as the
gap opens up, now with a steep slope, one is really starting an
avoided level crossing transition from a system that is still in its
ground state near the end of the plateau. Clearly, conclusions of
Refs.~\cite{DZ05a,CDDZ07a} apply.

While these preliminary insights are encouraging, more work is needed
to gain a complete picture.  In addition to the obvious intrinsic
interest of this problem there are applications that may benefit from
its thorough understanding. Let us mention one, slightly speculative:
{\it Adiabatic quantum computing} rests on the idea that a known
ground state of a simple Hamiltonian can be adiabatically transformed
into an initially unknown ground state that solves a problem encoded into the
structure of another Hamiltonian. The rate at which this transition
can be accomplished is limited by the size of the gap -- faster
transitions will result in errors appearing at the rate given by LZF,
Eq. (\ref{eq:LZF}). This will severely limit the speed of the
adiabatic computation -- its duration will increase exponentially with
the inverse of the size of the gap. But we have seen above that a gap can be widened,
and the rate of transitions can be brought down from exponential to
quadratic in $1/\tau_Q$. It remains to be seen whether (and how) these
changes can be used to suggest improvements in adiabatic quantum
computing, but the preliminary results we have reached here are
certainly suggestive.

This research was supported by the LDRD program funded by DoE at the
Los Alamos National Laboratory. The research of UD was supported by a
Marie Curie Intra-European Fellowship within the 6th European
Community Framework Programme and by the EPSRC (UK) through the QIP
IRC (GR/S82176/01) and EuroQUAM project EP/E041612/1.


\begin{thebibliography}{99}
\bibitem{ZZ} It is not yet clear however, whether these seeds are sown before or after 
the transition. There is numerical evidence that in thermodynamic transitions the instant
$\hat t$ after the quench plays a dominant role \cite{Gandra}. On the other hand, in quantum phase 
transitions both instants play a similar role \cite{DZ05a}, so that the effective power laws that follow
are a compromise (usually ``won" by the steeper dependence). Fortunately, when critical
exponents are the same on both sides of the transition -- as is the case in this paper --
this distinction is of no importance.
\end{thebibliography}
\end{document}